\documentclass[12pt]{article}
\title{Parameterized summation relations for the Stieltjes constants}
\author{Mark W. Coffey\\
Department of Physics\\
Colorado School of Mines\\
Golden, CO  80401\\
(Received $\mbox{~~~~~~~~~~~~~~~~~~~~~~~~~~~~~~~2010}$)}
\date{June 12, 2010}
\begin{document}
\maketitle
\baselineskip=25 pt
\begin{abstract}

The Stieltjes constants $\gamma_k(a)$ appear in the regular part of the Laurent expansion of the Hurwitz zeta function about its only polar singularity at $s=1$.
We present multi-parameter summation relations for these constants that result 
from identities for the Hurwitz zeta function.  
We also present multi-parameter summation relations for functions $A_k(x)$ that
may be expressed as sums over the Stieltjes constants.
Integral representations, especially including Mellin transforms, play an
important role.  As a byproduct, reciprocity and other summatory relations for
polygamma functions and Bernoulli polynomials may be obtained.

\end{abstract}
 
\vspace{.25cm}
\baselineskip=15pt
\centerline{\bf Key words and phrases}
\medskip 

\noindent
Hurwitz zeta function, summation relation, Stieltjes constants, Mellin transform

\vfill
\centerline{\bf 2010 AMS codes} 
11M35, 11M06, 11Y60
 
\baselineskip=25pt
\pagebreak
\medskip
\centerline{\bf Statement of results}
\medskip

Let $\zeta(s,a)$ denote the Hurwitz zeta function and $\gamma_k(a)$ the Stieltjes constant.  We present multi-parameter summation relations for these constants.
For background on the Stieltjes constants, one may see 
\cite{briggs,ivic,kreminski,matsuoka,matsuoka2,mitrovic,stieltjes,wilton2}.  For
known summation relations \cite{coffeyjmaa,coffeystdiffs,coffeyprsa} may be consulted.
In the Laurent series about $s=1$,
$$\zeta(s,a)={1 \over {s-1}}+\sum_{k=0}^\infty {{(-1)^k} \over k!} \gamma_k(a)(s-1)^k, \eqno(1.1)$$
$\gamma_0(a)=-\psi(a)$, where $\psi=\Gamma'/\Gamma$ is the digamma function and 
$\Gamma$ is the Gamma function. 
For the coefficients corresponding to the Laurent expansion for the Riemann
zeta function $\zeta(s)$, one denotes $\gamma_k(1)=\gamma_k$.  The quantities 
$\gamma_k(a)$ are of interest in analytic number theory, asymptotic analyses, and other areas. 

We let ${\cal P}$ denote the set of prime numbers.

We have
{\newline \bf Proposition 1}.  Let $p \geq 1$ and $q \geq 1$ be integers, $b\geq 0$, 
and min$(p/q,q/p) >b$.  Then we have for integers $k \geq 0$
$$\sum_{r=1}^q \gamma_k\left({{pr} \over q}-b\right)=q(-1)^k {{\ln^{k+1}(q/p)} \over
{k+1}} $$
$$+{q \over p}\sum_{\ell=0}^{p-1} \sum_{j=0}^k (-1)^j {k \choose j} \ln^j\left({q \over
p}\right)\gamma_{k-j}\left[1+{{(\ell-b)q} \over p}\right].  \eqno(1.2)$$

{\bf Proposition 2}.  Let $p \in {\cal P}$ and $m, N \geq 0$ be integers.  We have
$$(p-1)\gamma_m+{{\ln^{m+1}p} \over {m+1}}-\sum_{k=0}^{m-1}{m \choose k} \ln^{m-k} p
~\gamma_k$$
$$=(1-p) {{\ln^{m+1}p^{N+1}} \over {m+1}}+{1 \over p^N}\sum_{k=0}^m {m \choose k}
\ln^{m-k} p^{N+1} \sum_{\stackrel{1\leq j<p^{N+1}}{(j,p)=1}} \gamma_k\left({j \over
p^{N+1}}\right).  \eqno(1.3)$$

{\bf Proposition 3}.  Let $p \in {\cal P}$ and $m, N \geq 0$.  Then for any positive integer $k_p$ coprime to $p$ and nonnegative integers $\alpha$ and $\beta$ such that
$\alpha+jk_p=p \beta$ for some $j$ with $0 \leq j \leq p-1$, we have
$$(p-1)\gamma_m\left({\alpha \over k_p}\right)+{{\ln^{m+1}p} \over {m+1}}-\sum_{k=0}^{m-1}{m \choose k} \ln^{m-k} p
~\gamma_k\left({\beta \over k_p}\right)$$
$$=(1-p) {{\ln^{m+1}p^{N+1}} \over {m+1}}+{1 \over p^N}\sum_{k=0}^m {m \choose k}
\ln^{m-k} p^{N+1} \sum_{\stackrel{j \equiv \alpha (\mbox{\tiny{mod}} ~k_p)}{(j,p)=1}} \gamma_k\left({j \over {k_p p^{N+1}}}\right).  \eqno(1.4)$$
On the right side, the sum is over all integers $j$ of the form $\alpha+k_p \ell$ and
$\alpha \leq j <\alpha +k_p p^{N+1}$.

From these Propositions we obtain many Corollaries.  As an illustration we give some
of those resulting from Proposition 1.  For this purpose, we let $B_n(x)$ be the
Bernoulli polynomial of degree $n$ (e.g., \cite{nbs}, Ch. 23.1) and $\psi^{(j)}$ the
polygamma function (e.g., \cite{nbs}, Ch. 6.4).  We have the following.

{\bf Corollary 1}.  For $p,q \geq 1$ integers, $b\geq 0$, and min$(p/q,q/p) >b$, we
have
$$\ln q +{1 \over q}\sum_{r=0}^{q-1}\psi\left({{pr} \over q}-b\right)=\ln p+{1 \over p}
\sum_{\ell=0}^{p-1}\psi\left[(\ell-b){q \over p}\right].  \eqno(1.5)$$

{\bf Corollary 2}.  For $p,q \geq 1$, $n>1$ integers, $b\geq 0$, and min$(p/q,q/p) >b$, we have
$${1 \over q}\sum_{r=0}^{q-1}\psi^{(n-1)}\left({{pr} \over q}-b\right)={1 \over p}
\left({q \over p}\right)^{n-1}\sum_{\ell=0}^{p-1}\psi^{(n-1)}\left[(\ell-b){q \over p}\right].  \eqno(1.6)$$

{\bf Corollary 3}. For $p,q \geq 1$, $m \geq 0$ integers, $b\geq 0$, and min$(p/q,q/p) >b$, we have
$$\sum_{r=1}^q B_m\left({{pr} \over q}-b\right)=\left({q \over p}\right)^{1-m}
\sum_{\ell=0}^{p-1} B_m\left[1+(\ell-b){q \over p}\right].  \eqno(1.7)$$

The functions for integers $k$ 
$$A_k(q) \equiv k \left.{\partial \over {\partial z}}\zeta(z,q)\right|_{z=1-k},
\eqno(1.8)$$
are very useful in evaluating integrals over the Hurwitz zeta function
\cite{moll}.  They may be written in terms of the Stieltjes constants as
$$A_k(q)=-{1 \over k}-k\sum_{n=0}^\infty {{\gamma_{n+1}(q)} \over {n!}}k^n. 
\eqno(1.9)$$
We present representative summation relations for the functions $A_k(q)$.  We have
{\newline \bf Proposition 4}.  Let $p \geq 1$ and $q \geq 1$ be integers, $b\geq 0$, 
and min$(p/q,q/p) >b$.  Then we have
$$\sum_{r=1}^q \left[A_k\left({{pr} \over q}-b\right)+\ln\left({q \over p}
\right)B_k\left({{pr} \over q}-b\right)\right]=\left({q \over p}\right)^{1-k}
\sum_{\ell=0}^{p-1} A_k\left(1+{{(\ell-b)q} \over p}\right).  \eqno(1.10)$$

{\bf Proposition 5}.  For $p \in {\cal P}$ and integer $N \geq 0$ we have
$$(1-p^{k-1})A_k(1)-(-1)^k(\ln p)B_k[N(1-p^{k-1})+1]=
p^{(N+1)(k-1)}\sum_{\stackrel{1\leq j<p^{N+1}}{(j,p)=1}}A_k\left({j \over p^{N+1}}\right).  \eqno(1.11)$$
Here, $B_k=B_k(0)=(-1)^kB_k(1)$, $k \geq 0$ are the Bernoulli numbers.

\medskip
\centerline{\bf Proof of Propositions}

{\it Proposition 1}.  We apply Lemma 1 and then proceed as in the proof of Proposition 
5 of \cite{coffeyprsa}.

We first have the following.  
{\newline \bf Lemma 1}.  Let $p \geq 1$ and $q \geq 1$ be integers, $b\geq 0$, and
min$(p/q,q/p) >b$.  Then we have
$$\sum_{r=1}^q \zeta\left(s,{{pr} \over q}-b\right)=\left({q \over p}\right)^s ~
\sum_{\ell=0}^{p-1} \zeta\left(s,{{\ell q+p} \over p}-{{qb} \over p}\right).  \eqno(2.1)$$

{\it Proof}.  The Lemma may be proved in three different ways:  (i) interchange of a
double sum, (ii) use of an integral representation for the Hurwitz zeta function,
and (iii) evaluating the zeta functions associated with a certain rational function.
The Lemma is first demonstrated for Re $s>1$, and then by analytic continuation it
holds for all $s \in C$.

First method.  We have for Re $s>1$,
{\small{
$$\sum_{r=1}^q \zeta\left(s,{{pr} \over q}-b\right)=\sum_{r=1}^q\sum_{n=0}^\infty
{1 \over {\left(n+{{pr} \over q}-b\right)^s}}=\left({q \over p}\right)^s
\sum_{n=0}^\infty \left[\zeta \left(s,{{(n-b)q} \over p}+1\right)-\zeta \left(s,{{(n+p-b)q} \over p}+1\right) \right].  \eqno(2.2)$$}}
\noindent
Successive terms of the summand in blocks of length $p$ are then taken to find (2.1).
Indeed, (2.1) follows as the $u \to \infty$ limit of the relation

{\small{
$$\sum_{n=0}^u \left[\zeta \left(s,{{(n-b)q} \over p}+1\right)-\zeta \left(s,{{(n+p-b)q} \over p}+1\right) \right]$$
$$=\sum_{\ell=0}^{p-1} \left[\zeta\left(s,{{\ell q+p} \over p}-{{qb} \over p}\right)
-\zeta\left(s,{{uq +(\ell+1) q+p} \over p}-{{qb} \over p}\right)\right].  \eqno(2.3)$$}}

Second method.  We use a standard integral representation for Re $s>1$ and Re $a>0$,
$$\zeta(s,a)={1 \over {\Gamma(s)}}\int_0^\infty {{x^{s-1}e^{-(a-1)x}} \over {e^x-1}}
dx, \eqno(2.4)$$
to obtain
$$\sum_{r=1}^q \zeta\left(s,{{pr} \over q}-b\right)={1 \over {\Gamma(s)}}\int_0^\infty x^{s-1}{{e^{(b+1-p)x}(e^{px}-1)} \over {(e^x-1)(e^{px/q}-1)}}dx.  \eqno(2.5)$$
Similarly, we have
$$\left({q \over p}\right)^s \sum_{\ell=0}^{p-1} \zeta\left(s,{{\ell q+p} \over p}
-{{qb} \over p}\right)=\left({q \over p}\right)^s{1 \over {\Gamma(s)}}\int_0^\infty 
x^{s-1} {{(e^{qx}-1)e^{(b+1-p)qx/p}} \over {(e^x-1)(e^{qx/p}-1)}}dx$$
$$={1 \over {\Gamma(s)}}\int_0^\infty u^{s-1}{{e^{(b+1-p)u}(e^{pu}-1)} \over {(e^u-1)(e^{pu/q}-1)}}du,  \eqno(2.6)$$
where we used the scaling $u=qx/p$.

Third method.  We use the method of finding a zeta function $Z_f$ associated with a
rational function, (see the Appendix for a brief description), and express $Z_f$
in two different ways.  We employ the rational function
$$f(T)=\sum_{r=1}^q {T^{pr-bq} \over {1-T^q}}=\sum_{r=1}^q \sum_{n=0}^\infty
T^{pr+(n-b)q}.  \eqno(2.7)$$
By relation (A.3) of the Appendix, we have
$$Z_f(s)\Gamma(s)=\Gamma(s)q^{-s}\sum_{r=1}^q \zeta\left(s,{{pr} \over q}-b\right).
\eqno(2.8)$$
We also recognize that
$$f(T)=\sum_{\ell=0}^{p-1} {T^{p-bq} \over {1-T^p}}T^{\ell q}=\sum_{\ell=0}^{p-1}
T^p {T^{(\ell-b)q} \over {1-T^p}}.  \eqno(2.9)$$
The Mellin transform representation corresponding to this equation is given by
$$Z_f(s)\Gamma(s)={1 \over p^s}\sum_{\ell=0}^{p-1}\int_0^\infty v^{s-1} {e^{-(\ell-b)qv/p} \over {e^v-1}} dv
=p^{-s}\Gamma(s)\sum_{\ell=0}^{p-1} \zeta\left[s,1+(\ell-b){q \over p}\right].  \eqno(2.10)$$
Comparing (2.8) and (2.10), we have the Proposition.

Corollary 1 follows by putting $k=0$ in Proposition 1, and using the functional
equation of the digamma function $\psi(x+1)=\psi(x)+1/x$.  For instance, an
intermediate form is
$$\ln q +{1 \over q}\sum_{r=1}^{q}\psi\left({{pr} \over q}-b\right)=\ln p+{1 \over q}
\sum_{\ell=0}^{p-1}{1 \over {\ell-b}}+{1 \over p}\sum_{\ell=0}^{p-1}\psi\left[(\ell-b)
{q \over p}\right].  \eqno(2.11)$$

Corollary 2 follows by differentiating (2.11) with respect to $-b$ $(n-1)$ times.
Alternatively, it follows by taking $s=n$, $n>1$ an integer, in Lemma 1, using
the relation
$$\psi^{(n)}(x)=(-1)^{n+1}n!\zeta(n+1,x), \eqno(2.12)$$
and manipulating with the functional equation of the polygamma function,
$\psi^{(n-1)}(x+1)=\psi^{(n-1)}(x)+(-1)^{n-1}(n-1)!/x^n$.

Corollary 3 obtains from Lemma 1 since we have the relation $B_n(x)=-n\zeta(1-n,x)$
for $n>0$.
 
{\it Remarks}.  Equation (2.1) reduces properly at $p=1$.

In relation (2.1), there is cancellation of polar $q/(s-1)$ terms.

Of course, the $j=k$ term on the right side of (1.2) may be separated.

In \cite{coffeyprsa} (5.1) and (5.2), the summand factor $(-1)^k$ should read
$(-1)^j$.

We see that (2.6) corresponds to the following form of the function $f(T)$ used in
the third method:
$$f(T)=\sum_{n=0}^\infty T^p {{1-T^{pq}} \over {1-T^p}}T^{(n-b)q}={{T^{p-bq}(1-T^{pq})}
\over {(1-T^q)(1-T^p)}}.  \eqno(2.13)$$

Corollaries 1 and 2 correspond to successively differentiating Schobloch's
relatively little known reciprocity formula of 1884 for the function $\ln \Gamma$.
For a proof of this formula, see Theorem 3.7 in \cite{gksri}.

Owing to the many functional properties of the Bernoulli polynomials (e.g., 
\cite{nbs}, Ch. 23.1), such as
$$B_m\left(1-q {b \over p}\right)=(-1)^m B_m\left(q {b \over p}\right)
=(-1)^m q^{m-1} \sum_{k=0}^{q-1} B_m\left({b \over p}+{k \over q}\right), \eqno(2.14)$$
Corollary 3 may be written in many equivalent ways.  Corollary 3 is probably a new
reciprocity relation for the Bernoulli polynomials.

{\it Proposition 2}.  We use
{\newline \bf Lemma 2}.  For $p \in {\cal P}$ and $N \geq 0$, we have
$$(1-p^{-s})\zeta(s)=p^{-(N+1)s} \sum_{\stackrel{1\leq j<p^{N+1}}{(j,p)=1}} \zeta\left(s,{j \over p^{N+1}}\right).  \eqno(2.15)$$

Lemma 2 initially holds for Re $s>1$, and then by analytic continuation for all of C.
We have found that Lemma 2 and Lemma 3 below overlap with Proposition 1 in \cite{eie} 
as used in connection with a study of congruences of Bernoulli numbers.  As shown above for Lemma 1, there are alternative means of determining such identites.

{\it Proof}.  First method.  We use the method of finding a zeta function $Z_f$ associated with a rational function, (see the Appendix), expressing $Z_f$ in two different ways.  Associated with the function
$$f(T)={1 \over {1-T}}-{1 \over {1-T^p}}, \eqno(2.16)$$
we first have for Re $s>1$,
$$Z_f(s)=\sum_{n=1}^\infty {1 \over n^s}-\sum_{n=1}^\infty {1 \over {(pn)^s}}
=(1-p^{-s})\zeta(s).  \eqno(2.17)$$
Alternatively, we have
$$f(T)={{T+T^2+\ldots + T^{p-1}} \over {1-T^p}}={{(T+T^2+\ldots+ T^{p-1})(1+T^p +T^{2p}+\ldots + T^{p(p^N-1)})} \over {1-T^{p^{N+1}}}}$$
$$=\sum_{\stackrel{1\leq j<p^{N+1}}{(j,p)=1}} \sum_{k=0}^\infty T^{j+kp^{N+1}}.
\eqno(2.18)$$
Then for Re $s>1$,
$$Z_f(s)=p^{-(N+1)s} \sum_{\stackrel{1\leq j<p^{N+1}}{(j,p)=1}} \zeta\left(s,{j \over p^{N+1}}\right),  \eqno(2.19)$$
and the Lemma follows.

Second method for Lemma 2, using the integral representation (2.4).  We have
$$p^{-(N+1)s} \Gamma(s)\sum_{\stackrel{1\leq j<p^{N+1}}{(j,p)=1}}\zeta\left (s,{j 
\over p^{N+1}}\right)=p^{-(N+1)s}\sum_{\stackrel{1\leq j<p^{N+1}}{(j,p)=1}}
\int_0^\infty {t^{s-1} \over {e^t-1}}e^{-(j/p^{N+1}-1)t }dt$$
$$=p^{-(N+1)s} \sum_{k=1}^{p-1} \sum_{\ell=0}^{p^N-1} \int_0^\infty {t^{s-1} \over {e^t-1}}e^{-[(p\ell+k)/p^{N+1}-1]t }dt$$
$$=p^{-(N+1)s}\int_0^\infty t^{s-1} {{(e^{p^{-N}t}-e^{p^{-N-1}t})} \over {(e^{p^{-N-1}t}-1)(e^{p^{-N}t} -1)}}dt$$
$$=p^{-s}\int_0^\infty u^{s-1} {{(e^u-e^{u/p})} \over {(e^{u/p}-1)(e^u-1)}}du$$
$$=p^{-s}\int_0^\infty u^{s-1} \left[{1 \over {e^{u/p}-1}}-{1 \over {e^u-1}}\right]du$$
$$=\Gamma(s)(1-p^{-s})\zeta(s).  \eqno(2.20)$$

In Lemma 2, the polar term that cancels on each side is $(1-1/p)/(s-1)$.  This is
confirmed by the sum from the right side,
$$\sum_{\stackrel{1\leq j<p^{N+1}}{(j,p)=1}} 1=\varphi(p^{N+1})=p^{N+1}-p^N, \eqno(2.21)$$
where $\varphi$ is the Euler totient function.  By using the expansion (1.1), the
left side is 
$$(1-p^{-s})\zeta(s)=\left[1-{1 \over p}\sum_{j=0}^\infty {{(-1)^j} \over {j!}} \ln^j p
~(s-1)^j\right]\left[{1 \over {s-1}}+\sum_{k=0}^\infty {{(-1)^k} \over k!} \gamma_k(a)(s-1)^k\right],  \eqno(2.22)$$
and the right side is given by
$$p^{-(N+1)s}\sum_{\stackrel{1\leq j<p^{N+1}}{(j,p)=1}}\zeta\left({j \over p^{N+1}}\right)$$
$$={1 \over p^{N+1}}\sum_{\ell=0}^\infty {{(-1)^\ell} \over {\ell!}} \ln^\ell p^{N+1}
(s-1)^\ell \sum_{\stackrel{1\leq j<p^{N+1}}{(j,p)=1}}\left[{1 \over {s-1}}+\sum_{k=0}^\infty {{(-1)^k} \over k!} \gamma_k\left({j \over p^{N+1}}\right)(s-1)^k\right].  \eqno(2.23)$$
Manipulating the series in (2.22) and (2.23), identifying the coefficients of $(s-1)^m$
on both sides, and multiplying by $p$ gives Proposition 2.  

Remark.  From Proposition 2 with $m=0$ we have
{\newline \bf Corollary 4}.  For $p \in {\cal P}$ and $N \geq 0$ an integer, we have
$$(p-1)\gamma+\ln p=(1-p) \ln p^{N+1}-{1 \over p^N}\sum_{\stackrel{1\leq j<p^{N+1}}{(j,p)=1}} \psi\left({j \over p^{N+1}}\right). $$

{\it Proposition 3}.  We use
{\newline \bf Lemma 3}.  For $p \in {\cal P}$, $N\geq 0$, $\alpha\geq 0$, $\beta \geq 0$, and $(k_p,p)=1$ as in the Proposition, we have   
$$\zeta\left(s,{\alpha \over k_p}\right)-p^{-s}\zeta\left(s,{\beta \over k_p}\right)=p^{-(N+1)s} \sum_{\stackrel{j \equiv \alpha (\mbox{\tiny{mod}} ~k_p)}{(j,p)=1}} \zeta\left(s,{j \over {k_p p^{N+1}}}\right).  \eqno(2.24)$$

{\it Proof}.  We find zeta functions $Z_f$ associated with the rational function
$$f(T)={T^\alpha \over {1-T^{k_p}}}-{T^{p\beta} \over {1-T^{k_pp}}}.  \eqno(2.25)$$
We have the series for $|T|<1$,
$$f(T)=\sum_{n=0}^\infty T^{\alpha+k_pn}-\sum_{n=0}^\infty T^{p\beta+k_ppn}.  \eqno(2.26)$$
By (A.3) of the Appendix, we have
$$Z_f(s)\Gamma(s)=\int_0^\infty t^{s-1}f(e^{-t})dt=\Gamma(s)\left[k_p^{-s} \zeta \left
(s,{\alpha \over k_p}\right)-(k_pp)^{-s} \zeta \left (s,{\beta \over k_p}\right)\right].
\eqno(2.27)$$
Also expressing $f(T)$ as a rational function with denominator $1-T^{k_pp^{N+1}}$, we
find
$$f(T)=\sum_{\stackrel{j \equiv \alpha (\mbox{\tiny{mod}} ~k_p)}{(j,p)=1}} \sum_{n=0}^
\infty T^{j+nk_p p^{N+1}}.  \eqno(2.28)$$
Then we also have for Re $s>1$,
$$Z_f(s)\Gamma(s)=\int_0^\infty t^{s-1}f(e^{-t})dt=\Gamma(s)(k_pp^{N+1})^{-s} \sum_{\stackrel{j \equiv \alpha (\mbox{\tiny{mod}} ~k_p)}{(j,p)=1}} \zeta \left(s,{j \over {k_pp^{N+1}}}\right).   \eqno(2.29)$$
Equating (2.27) and (2.29), we obtain the Lemma.

The result of the Lemma extends to all of $C$ by analytic continuation.  There is
again cancellation of polar terms $(1-1/p)/(s-1)$.  Proceeding as in the proof of
Proposition 2, we obtain Proposition 3.

{\it Proposition 4}.  We first differentiate (2.1) with respect to $s$,
$$k\sum_{r=1}^q \zeta'\left(s,{{pr} \over q}-b\right)=\left({q \over p}\right)^s 
k \ln\left({q \over p}\right)~ \sum_{\ell=0}^{p-1} \zeta\left(s,{{\ell q+p} \over p}-{{qb} \over p}\right) 
+\left({q \over p}\right)^s k~ \sum_{\ell=0}^{p-1} \zeta'\left(s,{{\ell q+p} \over p}-{{qb} \over p}\right).  \eqno(2.30)$$    
We then put $s=1-k$ and use the definition (1.8), so that
$$\sum_{r=1}^q A_k \left({{pr} \over q}-b\right)=-\left({q \over p}\right)^{1-k} 
k \ln\left({q \over p}\right)~ \sum_{\ell=0}^{p-1} B_k\left({{\ell q+p} \over p}-{{qb} \over p}\right) +\left({q \over p}\right)^{1-k} k~ \sum_{\ell=0}^{p-1} A_k\left({{\ell q+p} \over p}-{{qb} \over p}\right).  \eqno(2.31)$$   
We now apply Corollary 3 for the Bernoulli polynomials to arrive at the Proposition.

{\it Proposition 5}.  By evaluating (2.15) at $s=1-k$ we have
\newline{\bf Corollary 5}.  We have
$$(-1)^k(1-p^{k-1})B_k=p^{(N+1)(k-1)} \sum_{\stackrel{1\leq j<p^{N+1}}{(j,p)=1}} B_k\left({j \over p^{N+1}}\right).  \eqno(2.32)$$
We now differentiate (2.15) with respect to $s$,
$$k(1-p^{-s})(\ln p)\zeta(s)+k(1-p^{-s})\zeta'(s)=-p^{-(N+1)s} (N+1)(\ln p) k\sum_{\stackrel{1\leq j<p^{N+1}}{(j,p)=1}} \zeta\left(s,{j \over p^{N+1}}\right)$$
$$+p^{-(N+1)s} k\sum_{\stackrel{1\leq j<p^{N+1}}{(j,p)=1}} \zeta'\left(s,
{j \over p^{N+1}}\right).  \eqno(2.33)$$
Putting $s=1-k$, using (1.8), we have
$$-p^{k-1}(\ln p)B_k(1)+(1-p^{k-1})A_k(1)=(N+1)p^{(N+1)(k-1)} (\ln p) \sum_{\stackrel{1\leq j<p^{N+1}}{(j,p)=1}} B_k\left({j \over p^{N+1}}\right)$$
$$+p^{(N+1)(k-1)} \sum_{\stackrel{1\leq j<p^{N+1}}{(j,p)=1}} A_k\left(
{j \over p^{N+1}}\right).  \eqno(2.34)$$
Applying Corollary 5 leads to (1.11).

\centerline{\bf Acknowledgement}

I thank H. Alzer for an inquiry that lead to Proposition 1.

\pagebreak
\centerline{\bf Appendix:  Zeta functions associated with a rational function}

Suppose that $p(T)$ is a polynomial in $T$ and $n_1$, $n_2$, ..., $n_r$ are 
positive integers.  We consider the rational function
$$f(T)={{p(T)} \over {(1-T^{n_1})(1-T^{n_2})\cdots(1-T^{n_r})}}.  \eqno(A.1)$$
We assume that about $T=0$ there is a power series expansion $f(T)=\sum_{n=0}^\infty 
a_n T^n$, $|T| < 1$.  Then the zeta function associated with $f(T)$ is given by
$$Z_f(s)=\sum_{n=1}^\infty {a_n \over n^s}, ~~~~\mbox{Re} ~s>r.  \eqno(A.2)$$
In addition, $Z_f$ is given by the Mellin transform
$$Z_f(s)={1 \over {\Gamma(s)}}\int_0^\infty t^{s-1}[f(e^{-t})-f(0)]dt.  \eqno(A.3)$$
So $Z_f$ is initially defined in a right half plane and has an analytic continuation.

\pagebreak

\end{document}